\documentclass[pre,twocolumn]{revtex4}
\usepackage{amssymb,amsmath,amsthm}
\usepackage[dvips]{graphicx}
\usepackage{verbatim}

\begin{document}
\title{BEYOND NETWORKS: OPINION FORMATION \\ IN TRIPLET-BASED POPULATIONS}
\author{DAMI\'AN H. ZANETTE}
\affiliation{Consejo Nacional de Investigaciones Cient\'{\i}ficas y
T\'ecnicas\\  Centro At\'omico Bariloche and Instituto Balseiro \\
8400 San Carlos de Bariloche, R\'{\i}o Negro, Argentina }

\begin{abstract}

\noindent We study a process of opinion formation in a population of
agents whose interaction pattern is defined on the basis of randomly
distributed groups of three agents, or triplets --in contrast to
networks, which are defined on the basis of agent pairs. Results for
the time needed to reach full consensus are compared between a
triplet-based structure with a given number of triplets and a random
network with the same number of triangles. The full-consensus time
in the triplet structure is systematically lower than in the
network. This discrepancy can be ascribed to differences in the
shape of the probability distribution for the number of triplets and
triangles per agent in each interaction pattern.
\bigskip


\end{abstract}

\maketitle

\noindent {\bf 1. Introduction} \smallskip

\noindent The representation of interaction patterns as graphs
–--or, as they are usually called, networks--– is a widespread
paradigm in science. Networks lie at the basis of the description of
a broad class of systems, ranging from intra-cellular molecular
reactions and neural tissues, to artificial objects such as the
internet [Newman {\it et al.}, 2006; Boccaletti {\it et al.}, 2006].
These systems are conceived as populations of active agents
(chemicals, neurons, computers), each of them occupying a network
node. A link between two agents represents the possibility that they
interact, mutually affecting their individual dynamics. Binary
interactions, in fact, are an implicit assumption within the network
picture. The inherent binary-like nature of all physical interaction
laws may explain why, from the viewpoint of physicists, do networks
provide such a satisfactory conceptual framework for applications of
statistical physics to other branches of knowledge.

\noindent  It is not always possible, however, to reduce the
dynamical laws of a multi-agent system to binary interactions.
Examples are found, for instance, in social dynamics [Starkey {\it
et al.}, 2000; Johnson, 2008]. Imagine a group of five people --–A,
E, I, O, and you--– who meet to discuss and make a decision on a
given issue. As soon as A exposes her views, your own opinion begins
to react to them, perhaps changing a little, or being reinforced.
Most probably, however, before your reaction to A's views is
definitively settled, you are already listening to E's. These have
been influenced by A's speech and, in turn, are expected to modify
both your preexisting opinion and the way it is affected by A's.  As
successive views are presented and overlap with each other, the
intricacy of their mutual effect grows. The group's final decision
will hardly be describable as the outcome of a sequence of binary
events, each involving just two people. In other words, from the
viewpoint of the group's dynamics, a representation in terms of a
mere collection of binary links joining its members seems
unsuitable. The intricate overlapping of the involved elementary
processes suggests that the group should be considered as a single
entity, subject to dynamical rules able to give an overall
description of the relation between the group's states and its
possible outcomes.

\noindent In this paper, we consider a process of opinion formation
in a population whose interaction pattern is defined in terms of
randomly distributed groups of three agents, or triplets. Emphasis
is put on comparison with the same process running on a random
Erd\H{o}s-R\'enyi  network, where the role of triplets is replaced
by triangles, each of them formed by three mutually linked nodes.
The Erd\H{o}s-R\'enyi network is built in such a way that the number
of triangles equals, on the average, the number of triplets in the
triplet-based structure. We find that the time needed to reach full
consensus in this structure is systematically shorter than in the
network, in spite of the fact that the number of triplets per agent
in the former is the same as the number of triangles per agents in
the latter. This leads us to compare in more detail the statistics
of triplets or triangles in each interaction pattern. The analysis
discloses the different effects of the two patterns on the dynamical
process taking place in the population.

\bigskip
\noindent {\bf 2. Multiplet-based populations} \smallskip

\noindent From the several possible ways in which a network can be
characterized [Newman {\it et al.}, 2006], the one which is most
suitable for the generalization we discuss in this paper is the
traditional definition given by graph theory [Gross \& Yellen,
1999]. A graph is defined as a set of vertices --or nodes, which we
associate with the members or agents in a population-- and a list of
edges --or links, which represent the potential interaction between
agents. Each edge is effectively defined by the pair of vertices
that it connects. In other words, for a given population, a network
is completely characterized by listing the pairs of nodes connected
by links. Over the population, thus, a network is a collection node
pairs.

\noindent This characterization of the structure of a population can
be immediately generalized. Instead of a collection of node pairs
(duplets) we specify a collection of node groups (multiplets), each
group containing a certain number $m$ of nodes ($m$-plet). In this
way, agents are aggregated into groups of different sizes. In
networks, an agent may participate of several interaction pairs,
depending on its connectivity. Similarly, in the mutiplet-based
population, a given agent can belong to more that one multiplet.
Multiplets are conceived to replace network links as the basic units
associated with the elementary interactions between agents [Johnson,
2008].

\begin{figure} [ht]
\begin{center}
\includegraphics
[width=\columnwidth,clip=true]{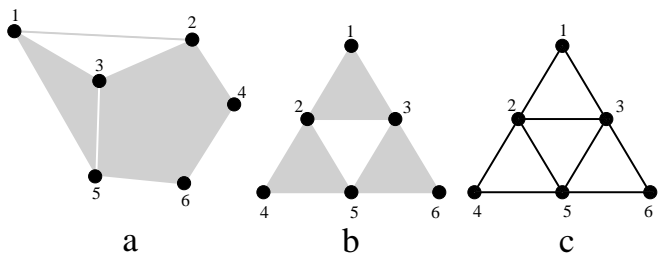} \caption{(a) A
$6$-agent population whose interaction pattern is given by a duplet,
a triplet, and a quintuplet. (b) A $6$-agent population, with three
triplets. (c) A $6$-node network, with $9$ links joining the nodes
which share triplets in (b).} \label{fig1}
\end{center}
\end{figure}

\noindent Figure \ref{fig1} gives two instances of small
multiplet-based populations. In Fig.~\ref{fig1}a, we have a
$6$-agent population whose interaction structure is defined by a
duplet, $(1,2)$, a triplet, $(1,3,5)$, and a quintuplet,
$(2,3,4,5,6)$. Agents $1$, $2$, $3$, and $5$ belong to more than one
multiplet. Figure \ref{fig1}b shows a $6$-agent population
structured into three triplets, $(1,2,3)$, $(2,4,5)$, and $(3,5,6)$.

\noindent Is there, however, a genuine difference between a
multiplet-based structure and a network? Wouldn't each multiplet be
equivalent to a fully connected group of vertices --a clique [Newman
{\it et al.}, 2006] or simplex [Johnson, 2008]-- in an ordinary
network? Comparison of Figs.~\ref{fig1}b and c shows, by means of a
simple example, that this is not the case. In Fig.~\ref{fig1}c, each
triplet of Fig. \ref{fig1}b has been replaced by the links which
interconnect all its agents. In the resulting network, however, the
triangle formed by agents $2$, $3$, and $5$, turns out to be
equivalent, from the viewpoint of its interaction pattern, to the
triangle formed, for instance, by $1$, $2$, and $3$. In the
structure of Fig. \ref{fig1}b, on the other hand, $(1,2,3)$ is
indeed one of the triplets in the population, while $(2,3,5)$ does
not exist as such.

\noindent A full appraisal of the implications of extending networks
to multiplet-based structures must in any case be sought through its
effects on the collective dynamics of the population. In the next
section, we consider a process of opinion formation [Krapivsky \&
Redner, 2003; Boccaletti {\it et al.}, 2006; Galam, 2008] and
compare results for the time needed to reach full consensus in
networks and in triplet-based structures. Differences can be
ascribed to rather nontrivial statistical properties of both kinds
of interaction pattern.

\bigskip
\noindent {\bf 3. Opinion dynamics in triplet structures}
\smallskip

\noindent The dynamical processes that may require a representation
of the population in terms of multiplets --for instance, the case of
decision making discussed in the Introduction-- need not be the same
as those which work on networks --such as, for instance,
peer-to-peer information transmission. For the sake of comparison
between the effects that these two different interaction patterns
have on dynamics, however, we analyze in this section the same
process of opinion formation running both on a network and on a
triplet-based structure.

\noindent Consider a population of $N$ agents where, at each time
$t$, the opinion $s_i(t)$ of each agent $i$ adopts one of two given
values, say, $s_i(t)=+1$ or $-1$. At each evolution step, three
agents $i$, $j$, and $k$, are chosen, and they adopt the opinion of
the majority among them, i. e.
\begin{eqnarray}
s_i (t+1) &=& s_j (t+1) = s_k (t+1) =  \nonumber \\ &=& {\rm sign}
[s_i(t)+s_j(t)+s_k(t)].
\end{eqnarray}
The choice of the three agents is performed according to the
structure of the population, as explained in the following.

\noindent Firstly, we consider an $N$-node Erd\H{o}s-R\'enyi random
network [Newman {\it et al.}, 2006], where each of the $N(N-1)/2$
possible links is effectively present with probability $p$. Thus,
for large $N$, the expected number of links is $L=pN^2/2$. In this
network, we define a triangle as a set of three mutually connected
nodes. The expected number of triangles is $\Delta =p^3N^3/6$. On
this network, the three agents chosen at each evolution step of the
process of opinion formation must form a triangle.

\noindent Secondly, we consider what could be called an
Erd\H{o}s-R\'enyi triplet-based structure, where each of the $N^3/6$
possible triplets in the $N$-agent population is effectively present
with probability $q$. Consequently, the expected number of triplets
is $T = q N^3 / 6$. In the process of opinion formation running on
this structure, the three agents chosen at each step must belong to
the same triplet. Taking $q=p^3$, we have on the average $T=\Delta$.
With this choice, the dynamics on the network and on the triplet
structure can be quantitatively compared.

\noindent We study the process of opinion formation on these two
interaction patterns by means of numerical simulations. We focus on
the time needed to reach full consensus, when all the agents share
the same opinion. Three stages contribute to the full-consensus time
in a finite population. If in the initial condition the two opinions
are more or less balanced over the population, the first stage
corresponds to the escape from the vicinity of the state where
opinions are equally frequent, to a region where the unbalance
between opinions is of order $N$. In this stage, the dynamics is
equivalent to a symmetric random walk. Therefore, its duration is of
order $N^2$. In the second stage --whose duration is of order $N$--
the dynamics is essentially deterministic, and corresponds to the
increasing prevalence of the majority opinion over the dissenters.
Finally, when only a few dissenters remain, the dynamics is
stochastically driven by the now-infrequent events where dissenters
are chosen and converted to the majority opinion. The duration of
this final state is also of order $N$. In our numerical simulations,
we skip the first stage by taking a number of dissenters equal to
$N/4$, so that $75 \%$ of the population is already consensual. The
initial dissenters are distributed at random.

\noindent The numerical construction of the two interaction patterns
is as follows. We first fix the number $T$ of triplets, and choose
at random the corresponding $T$ groups of three agents. Two or more
triplets containing the same three agents are avoided. With this
procedure, the triplet-based structure is completely defined. Then,
we construct a random network with $L=N(3T /4)^{1/3}$ links.
Multiple links between any two agents are forbidden. The expected
number of triangles for that number of links is, precisely, $T$. For
a given realization of the random network with $L$ links, the number
of triangles $\Delta$ --which is close, but not necessarily equal,
to $T$-- is determined by simple counting. The full-consensus time
$t_{\rm c}$, measured in evolution steps, is determined as a
function of $\Delta$ in the network, and of $T$ in the triplet
structure. Results for fixed networks and triplet structures, with
given values of $\Delta$ and $T$, are averaged over $ 10^4$
realizations for different initial opinion distributions.

\begin{figure} [ht]
\begin{center}
\includegraphics[width=\columnwidth,clip=true]{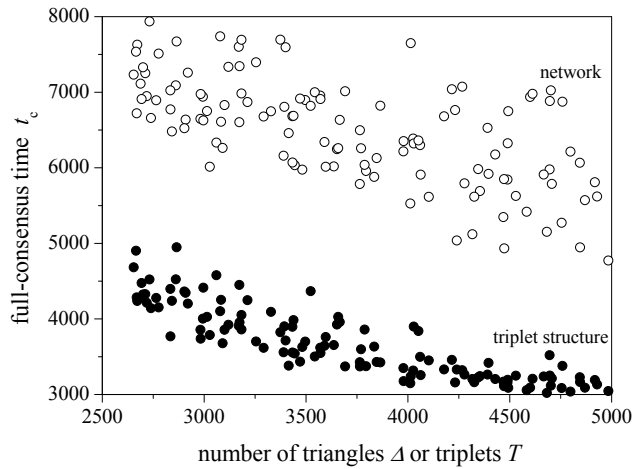}
\caption{Numerical results for the time needed to reach full
consensus, $t_{\rm c}$, measured in evolution steps, for a process
of opinion formation running on a population of $N=1000$ agents
whose interaction pattern is a network (empty dots) or a
triplet-based structure (full dots). The full-consensus time $t_{\rm
c}$ is plotted against the number of triangles $\Delta$ or triplets
$T$ in each pattern. Each dot stands for an average over $10^4$
realizations of the initial opinion distribution.} \label{fig2}
\end{center}
\end{figure}

\noindent Figure \ref{fig2} shows $t_{\rm c}$ as a function of
$\Delta$ and $T$ for a population of $N=1000$ agents. Empty and full
dots respectively correspond to results for the network ($t_{\rm c}$
vs. $\Delta$) and the triplet structure ($t_{\rm c}$ vs. $T$), for
the number of triangles and triplets ranging from $2500$ to $5000$.
We find that the number of evolution steps needed to reach consensus
on the network is more disperse and systematically larger than in
the triplet-based structure. In the range displayed, the two values
of $t_{\rm c}$ differ by a factor of almost two. For larger values
of $\Delta$ and $T$, as it may be expected, the full-consensus times
for the two interaction patterns approach each other. In fact, in
the limit where all the possible links and all the possible triplets
are present, $\Delta = T= N^3/6$, the two times must be identical.
Conversely, they are increasingly different as the interaction
patterns become more sparse. Consistent results were obtained for
systems of various sizes.

\noindent What is the origin of the difference between the
full-consensus times in the network and in the triplet structure? In
view of the fact that the dynamical rules of opinion formation are
identical for the two populations, the answer is to be found in
their interaction patterns. In the next section, we provide evidence
that an excess of network nodes belonging to a small number of
triangles with respect to the agents belonging to the same number of
triplets explains the difference. This requires to analyze in
detail, for each kind of pattern, the distribution of triangles or
triplets over the population.

\bigskip
\noindent {\bf 4. Statistics of triangles and triplets}
\smallskip

\noindent In our numerical simulations, we have compared the
full-consensus times in a network and in a triplet structure
constructed in such a way that the number $\Delta$ of triangles in
the former is, on the average, the same as the number $T$ of
triplets in the latter. This implies, in particular, that the
average number of triangles per agent, $3\Delta /N$, equals the
average number of triplets per agent, $3T /N$. In the following we
show that these two coincident averages correspond however to
different distributions.  More specifically, the fraction of agents
which belong to a certain number of triangles in the network is in
general different to the fraction of agents which belong to a
certain number of triplets in the triplet-based structure.

\noindent The fraction of nodes which belong to exactly $\delta$
triangles in an $N$-node Erd\H{o}s-R\'enyi network can be evaluated
by first considering the probability $P_n$ that a generic node is
connected to exactly $n$ of the possible $N-1$ neighbors, which is
given by the binomial distribution
\begin{equation}
P_n = \left(
\begin{array}{c} N-1 \\ n \end{array}
\right) p^n (1-p)^{N-1-n} ,
\end{equation}
where $p$ is the probability that a network link is effectively
present (see Section 3). The $n$ neighbors form $n(n-1)/2$ pairs,
some of which can in turn be joined by links, thus forming
triangles. The probability $P_\delta (n)$ that the node with $n$
neighbors belongs to $\delta$ of those triangles reads
\begin{equation}
P_\delta (n) = \left(
\begin{array}{c} n(n-1)/2 \\ \delta \end{array}
\right) p^\delta (1-p)^{n(n-1)/2-\delta} .
\end{equation}
The total probability $P_\delta$ that, irrespectively of the number
of neighbors, a node belongs to exactly $\delta$ triangles is given
by the total contribution over the different values of $n$:
\begin{equation} \label{Pdelta}
P_\delta = \sum_{n=0}^{N-1} P_n P_\delta(n).
\end{equation}
Some little algebra shows that, as advanced, the average number of
triangles per agent is, in the limit of large $N$, $\langle \delta
\rangle=p^3 N^2/2=3\Delta /N$. The width of the distribution can be
computed by approximating $P_n$ by a Gaussian, yielding for the
variance
\begin{equation}
\sigma_\delta^2 = \frac{N}{4} p^3 (1-p) [4N^2p^2-2N(p^2+p-1)-1].
\end{equation}
Which term dominates this expression for large $N$ depends on how
$p$ is assumed to depend on $N$ in that limit, as discussed below.

\noindent The probability  distribution $P_\tau$ of the number of
triplets per agent $\tau$ in the $N$-agent population with $T$
triplets is found, more straightforwardly, to be the binomial
\begin{equation} \label{Ptau}
P_\tau = \left(
\begin{array}{c} T \\  \tau  \end{array}
\right) \left(  \frac{3}{N} \right)^\tau  \left( 1- \frac{3}{N}
\right)^{T-  \tau } .
\end{equation}
The mean value, as predicted, is $\langle  \tau  \rangle=3 T /N$.
For large $N$,  the variance is $ \sigma_\tau^2  = 3T/N $.

\begin{figure} [ht]
\begin{center}
\includegraphics[width=\columnwidth,clip=true]{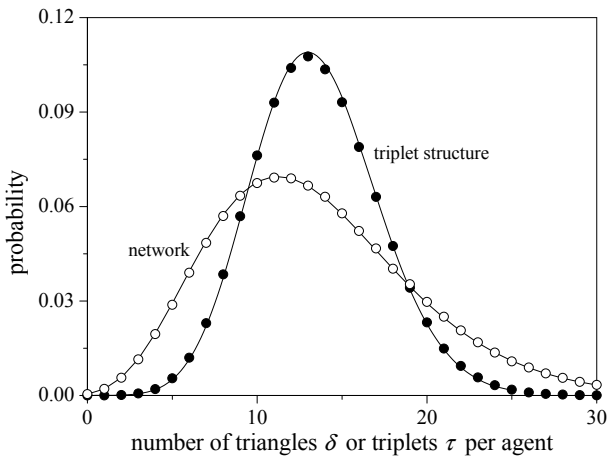}
\caption{Distribution probability for the number of triangles per
agent in a network (empty dots) and the number of triplets per agent
in a triplet-based structure (full dots), measured numerically for a
population of $N=1000$ agents with $4500$ triangles or triplets.
Curves stand for the analytical results of Eqs. (\ref{Pdelta}) and
(\ref{Ptau}).} \label{fig3}
\end{center}
\end{figure}

\noindent Figure \ref{fig3} shows the probability distributions
$P_\delta$ and $P_\tau$ for a population of $N=1000$ agents with
$4500$ triangles or triplets. Curves stand for the analytical
expressions of Eqs. (\ref{Pdelta}) and (\ref{Ptau}), and dots
correspond to the numerical evaluation of the probabilities, which
turn out to be in full agreement with the analytical results. The
average number of triangles or triplets per agent is the same for
both distributions, $\langle \delta \rangle=\langle \tau \rangle =
13.5$. Regarding their shapes and widths, on the other hand, the two
functions are substantially different. For the triplet structure, we
recognize the symmetric binomial distribution of Eq. (\ref{Ptau}).
The distribution of triangles per agent, in contrast, is clearly
asymmetric, with a rather long tail for large $\delta$. Since,
moreover, the maximum of $P_\delta$ is sensibly lower than that of
$P_\tau$, for small values of  $\delta$ and $\tau$ the probability
of triangles per agent is higher than that of triplets per agent. In
other words, the fraction of agents which belong to a small number
of triangles in the network is larger than the fraction of agents
belonging to the same number of triplets in the triplet-based
structure.

\noindent The difference between $P_\delta$ and $P_\tau$ for small
values of their variables provides an explanation for our
observation in Section 3 that the full-consensus time in the network
is systematically larger than in the triplet structure. Once the
quasi-deterministic stage of the opinion formation process has
elapsed, only a few dissenters remain in the population. These last
non-consensual agents are expected to belong, on the average, to a
small number of triangles or triplets. In fact, those agents
belonging to a relatively large number of triangles or triples have
had more opportunities of being selected during the preceding stage,
thus reaching consensus earlier. For a given number of agents in the
left end of the distributions $P_\delta$ and $P_\tau$, however, the
average number of triplets per agent is larger than the
corresponding number of triangles. Consequently, the last dissenters
have a larger probability per evolution step to be chosen, and their
opinion thus changed, in the triplet-based structure. The final
evolution stage will therefore elapse faster than in the network.

\noindent We end our discussion on the distributions $P_\delta$ and
$P_\tau$ by comparing their limits for large populations, $N\to
\infty$. As advanced above, to effectively perform the limit it is
necessary to specify how the interaction pattern changes as $N$
grows. We consider three cases: (i) constant total number of
triangles or triplets; (ii) constant average number of triangles or
triplets per agent; and (iii) constant fraction of effectively
present triangles or triplets. In case (i), $\Delta$ and $T$ are
independent of $N$, and the probability $p$ of the Erd\H{o}s-R\'enyi
network scales as $p \sim N^{-1}$. In case (ii), both $\Delta$ and
$T$ must be proportional to $N$, while $p \sim N^{-2/3}$. In case
(iii), $\Delta$ and $T$ scale as $N^3$, and $p$ remains constant.

\noindent It is clear that, since in case (i) the number of
triangles or triplets is kept constant as the population grows, the
distributions $P_\delta$ and $P_\tau$ in the limit of large $N$ are
trivial. They both collapse to delta functions, with vanishing mean
value and variance. Case (ii) is more interesting. In this case, as
$N$ grows, the mean values $\langle  \delta  \rangle$ and $\langle
\tau \rangle$, and the variances $\sigma_\delta^2$ and
$\sigma_\tau^2$ approach constant values. By construction, the
limiting values of $\langle  \delta  \rangle$ and $\langle  \tau
\rangle$ are identical. Also the two variances turn out to be
identical in the limit. Interestingly, however, their mutual
approaching for large $N$ is very slow, with
\begin{equation}
\sigma_\delta^2 \approx \sigma_\tau^2 \left( 1+ k N^{-1/3} \right)
\end{equation}
where $k$ is a constant. This suggests that the dynamical effects of
the difference between $P_\delta$ and $P_\tau$ --such as the
different duration of the process of opinion formation discussed in
Section 3-- may be perceivable even in very large populations.
Finally, in case (iii) the limiting distributions are drastically
different. In fact, for the network the variance grows as
$\sigma_\tau^2 \sim N^5$, while for the triplet structure we find
$\sigma_\delta^2 \sim N^2$.

\bigskip
\noindent {\bf 5. Conclusion}
\smallskip

\noindent Multiplet-based structures are expected to replace
networks as a representation of the interaction pattern underlying a
population when the relevant dynamical processes cannot be
assimilated to sequential binary interactions. This paper, however,
has been focused on a comparison of the dynamical effects of the two
kinds of pattern on the same process of opinion formation, driven by
a majority rule in groups of three agents. By means of numerical
simulations, we have determined the time needed to reach full
consensus, on one hand, in a population structured on the basis of
randomly distributed triplets and, on the other, on a population of
the same size whose interactions are described by a random
Erd\H{o}s-R\'enyi network forming as many triangles as the number of
triplets in the triplet-based structure. In spite of the strong
similarity between the two versions of the opinion formation
process, there is a noticeable difference in the full-consensus time
measured on both interaction patterns. We have shown that the
difference can be ascribed to higher-order statistical properties in
the distribution of triangles or triplets per agent in each pattern.
This conclusion points out the significant effect of the kind of
underlying interaction pattern in the collective dynamics of the
population.

\noindent The present results should be considered as a preliminary
step in the study of the dynamical effects of multiplet-based
structures. Further work may address dynamical processes more
specific to this kind of structure, for which networks do not
provide a suitable representation of interactions. The possibility
that in a given structure the size of multiplets in not homogeneous,
but varies according to a prescribed distribution; the extension of
the network-specific notion of degree distribution to multiplets;
the consideration of small-world or scale-free multiplet structures
are, among other issues, worth investigating.

\bigskip

\noindent {\bf Acknowledgments} \smallskip

\noindent The author acknowledges financial support from grants PIP
5114 (CONICET, Argentina) and PICT 17/943/04 (ANPCyT, Argentina).

\bigskip

\noindent {\bf References} \smallskip

\noindent Boccaletti, S., Latora, V.,  Moreno, Y., Chavez, M. \&
Hwanga, D.-U. [2006] ``Complex networks: Structure and dynamics,''
{\it Phys. Rep.} {\bf 424}, 175-308.

\noindent Galam, S. [2008] ``Sociophysics and the Forming of Public
Opinion: Threshold versus Non Threshold Dynamics,''
arXiv:0803.2453v1 [physics.soc-ph].

\noindent Gross, J. \& Yellen, J. [1999] {\it Graph Theory and
Applications} (CRC Press, Boca Raton).

\noindent Johnson, J. [2008] ``Multidimensional Events in Multilevel
Systems,'' in Albeverio, S., Andrey, D., Giordano, P. \& Vancheri,
A., eds.  {\it The Dynamics of Complex Urban Systems}
(Physica-Verlag, Heidelberg).

\noindent Krapivsky, P. \& Redner, S. [2003] `` Dynamics of Majority
Rule in Two-State Interacting Spin Systems,'' {\it Phys. Rev. Lett.}
{\bf 90}, 238701.

\noindent Newman, M.,  Barabasi, A.-L. \&  Watts, D. J., eds. [2006]
{\it The Structure and Dynamics of Networks} (Princeton University
Press, Princeton).

\noindent Starkey, K.,  Barnatt, Ch. \&  Tempest, S. [2000] ``Beyond
Networks and Hierarchies: Latent Organisation in the UK Television
Industry,'' {\it Organization Science} {\bf 11}, 299-305.

\end{document}